\documentclass[onecolumn,english,aps,prl]{revtex4}
\usepackage[T1]{fontenc}

\usepackage{amsmath,graphicx,amssymb,epsfig,babel,dsfont,float}

\begin{document}

\title{Smart thermal management with near-field thermal radiation}

\author{Ivan Latella}
\affiliation{Departament de F\'{i}sica de la Mat\`{e}ria Condensada, Universitat de Barcelona, Mart\'{i} i Franqu\`{e}s 1, 08028
Barcelona, Spain.}

\author{Svend-Age Biehs}
\affiliation{Institut f\"{u}r Physik, Carl von Ossietzky Universit\"{a}t, D-26111 Oldenburg, Germany.}

\author{Philippe Ben-Abdallah}
\email{pba@institutoptique.fr} 
\affiliation{Laboratoire Charles Fabry, UMR 8501, Institut d'Optique, CNRS, Universit\'{e} Paris-Saclay, 2 Avenue Augustin Fresnel, 91127 Palaiseau Cedex, France.}

\date{\today}



\begin{abstract}
When two objects at different temperatures are separated by a vacuum gap they can exchange heat by radiation only. At large separation distances (far-field regime) the amount of transferred heat flux is limited by Stefan-Boltzmann's law (blackbody limit). In contrast, at subwavelength distances (near-field regime) this limit can be exceeded by orders of magnitude thanks to the contributions of evanescent waves. This article reviews the recent progress on the passive and active control of near-field radiative heat exchange in two- and many-body systems.
\end{abstract}

\maketitle

\section{Introduction}

The control of electron flow in solids is at the origin of modern electronics which has revolutionized our daily life. The diode and the transistor introduced by Braun~\cite{Braun} and Bardeen~\cite{Bardeen}, respectively are undoubtedly the cornerstones of modern information technologies. Such devices allow for rectifying, switching, modulating and even amplifying the electric currents. Astonishingly, until very recently no thermal analogues of these building blocks were devised to exert a similar control on the heat flux. An important step forward in this direction has nevertheless been carried out by Baowen Li and co-workers~\cite{Li2004,Li2006} and by Chang et al.~\cite{Chang} at the beginning of 2000’s, when they proposed a phononic counterpart of the diode and transistor~\cite{Li2012}. These pioneer works have paved the way to a technology, also 
called ``thermotronics'' in analogy to traditional electronics, where electrical currents and voltage biases are replaced by heat currents and temperature biases to control heat conduction though a network of solid elements. A recent review~\cite{Li2021} summarizes the last developments carried out to control heat flux carried conduction at both macroscale and microscale using artificial structures.

However, heat transport mediated by phonons in solid networks suffers from some weaknesses of fundamental nature which intrinsically limit the performance of this technology. One of these limitations is linked to the speed of acoustic phonons itself (the speed of sound) which bounds the operational speed of these devices. Another intrinsic limitation of phononic devices is the presence of local Kapitza resistances which come from the mismatch of vibrational modes supported by the different solid elements in the network. This resistance can drastically reduce the heat transported across the system. To overcome these limitations, concepts for a purely photonic technology have been proposed as an alternative way to handle heat transfer at the nanoscale. 
In the present work, we review recent developments carried out in this direction. After briefly introducing the theoretical framework commonly used to described the radiative heat transfer in the near-field regime between two or several solid bodies, we describe the main physical mechanisms and related device concepts which allow for a passive and active control of radiative heat transfer at the nanoscale. Finally, we conclude this review by suggesting future research directions for advanced thermal management with thermal photons. 

\section{Some basics on the near-field heat transfer}

\begin{figure}
	\centering
	\includegraphics[scale=1]{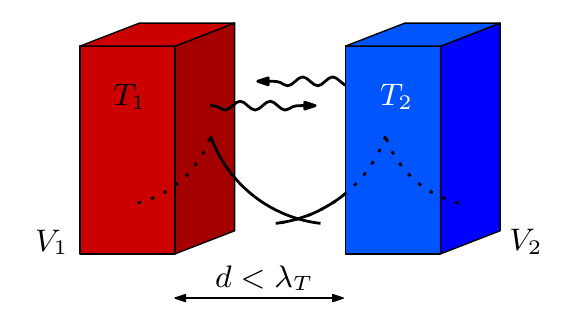}
	\caption{Sketch of radiative heat exchanges between two solids of volumes $V_1$ and $V_2$ held at temperatures $T_1$ and $T_2$ and separated by a vacuum gap of thickness $d$. At large separation distances, heat exchanges are mediated by propagating photons (wavy arrows). At subwavelength distances ($d<\lambda_T$, $\lambda_T$ being the thermal wavelength), the heat transfer is enhanced by the the contribution of evanescent waves localized on the surface of bodies.} 
	\label{2_body}
\end{figure}

The radiative heat transfer between distant objects, in the far field, is bounded by the blackbody limit given by the Stefan-Boltzmann law~\cite{Planck}. The transport of heat in this situation is mediated by propagating modes of the electromagnetic radiation emitted by the objects. 
When separation distances are smaller than the thermal wavelength $\lambda_T$ defined by Wien’s displacement law, which is about $10\,\mu$m at room temperature, near-field effects become relevant due to the contribution of evanescent modes of the electromagnetic field confined close to the surface of the objects. By bringing them at separations $d<\lambda_T$, the blackbody limit can notably be overcome owing to this near-field contribution from evanescent waves~\cite{Cravalho,Polder,Pendry,Joulain2005,Volokitin}. Hence, the radiative heat flux exchanged in near-field between two silica samples separated by a distance $d=100\:nm$ around the ambient temperature with a temperature gradient $\Delta T=50\:K$ is $\phi\simeq 20 000\:W.m^{-2}$, while the blackbody limit is $\phi_{BB}=\sigma T^3\Delta T\simeq75\:W.m^{-2}$ and the solar flux used for conventional photovoltaics is about $\phi_S=1000\:W.m^{-2}$, $\sigma$ being the Stefan-Boltzmann constant.

The near-field radiative heat exchange in a given configuration of several solid objects in a thermal non-equilibrium situation are commonly calculated in the framework of fluctuational electrodynamics. To illustrate the basic principles of this approach, let us consider the simple example of two objects with volumes $V_1$ and $V_2$ held at temperatures $T_1$ and $T_2$  which are separated by a vacuum gap of thickness $d$ as sketched in Fig.~\ref{2_body}. The thermal motion of charges within each of these objects induce fluctuational current densities  ${\bf j}_i({\bf r},\omega)$ ($i = 1,2$) which themselves induce fluctational electric and magnetic fields ${\bf E}_i$ and ${\bf H}_i$ fulfilling the stochastic Maxwell equations~\cite{Rytov} 
\begin{eqnarray}
  \nabla \times {\bf E}_i({\bf r},\omega) & = & i \omega \mu_0 {\bf H}_i({\bf r},\omega), \\
  \nabla \times {\bf H}_i({\bf r},\omega) & = & -i \omega \epsilon_0 \mathbf{\epsilon}_i({\bf r},\omega) 
  {\bf E}_i ({\bf r},\omega) +  {\bf j}_i({\bf r},\omega) ,
\end{eqnarray}
where $\epsilon_i({\bf r},\omega)$ denotes the local dielectric tensor of object $i$ (here assumed to be non-magnetic) at point $\bf r$; $\epsilon_0$ and $\mu_0$ are the permittivity and permeability of vacuum. The linearity of these equations allows us to relate ${\bf E}_i$ and ${\bf H}_i$ to the source currents ${\bf j}_i ({\bf r},\omega)$ as follows~\cite{Rytov}
\begin{align}
	{\bf E}_i ({\bf r},\omega) &=  i \omega \mu_0 \int_{V_i} {\rm d}^3 r' \, \mathds{G}^{EE}({\bf r},{\bf r'}) {\bf j}_i({\bf r'},\omega), \\
	{\bf H}_i ({\bf r},\omega) &=  i \omega \mu_0 \int_{V_i} {\rm d}^3 r' \,\mathds{G}^{HE}({\bf r},{\bf r'}) {\bf j}_i({\bf r'},\omega),
\end{align}
where $\mathds{G}^{EE}$ and $\mathds{G}^{HE}$ denote the linear electric and magnetic response tensors also called the dyadic Green functions of the system. From these expressions one can determine the mean Poynting vector
\begin{equation}
	\langle \boldsymbol{\Pi}_i ({\bf r},\omega)\rangle = 2 \mathcal{R}e\langle{\bf E}_i \times{\bf H}^*_i \rangle
\end{equation}
which can be readily expressed in terms of both the Green tensor components and the correlations functions of fluctuating currents. 
Assuming that the objects are in local thermal equilibrium, then according to the fluctuation-dissipation theorem these correlations 
are related to the local temperature by the following expression~\cite{Eckhardt}
\begin{equation}
	\langle j_{i,\mu} ({\bf r},\omega) j_{i,\nu}^{\ast}({\bf r}^{\prime},\omega) \rangle =\frac{2 \hslash \omega^2 \epsilon_0}{\pi} [\epsilon_{i, \mu \nu}({\bf r},\omega)-\epsilon^*_{i, \nu \mu}({\bf r},\omega)] n(\omega,T_i) \delta({\bf r}-{\bf r}^{\prime}) ,
\end{equation}
where $\hslash$ is the Planck constant and $n(\omega,T_i) = 1/(\exp[\hslash \omega/k_{\rm B}T_i]-1)$ is the Bose-Einstein distribution function at temperature $T_i$; $k_{\rm B}$ is the Boltzmann constant. It follows that the 
Poynting vector can be expressed in terms of all local temperatures inside the system.
The spectral radiative power $P_{1\leftrightarrow 2}(\omega)$ exchanged between the two objects can be obtained by integrating the flux expressed by the Poynting vector over the surfaces $A_i = \partial V_i$ of two bodies as follows
\begin{equation}
	P_{1\leftrightarrow 2}(\omega)=\int_{{A_2}}{\bf dA_2}\cdot \langle\boldsymbol{\Pi}_1({\bf r},\omega)\rangle - \int_{{A_1}} {\bf dA_1}\cdot\langle \boldsymbol{\Pi}_2({\bf r},\omega)\rangle.
\end{equation}
Note, that here this expression is only valid as long as the source currents in both objects are uncorrelated~\cite{Rytov}.
Finally, the net power exchanged between the two bodies is obtained by summation over all frequencies (i.e. $\mathcal{P}_{net}=\int_0^\infty \frac{d\omega}{2\pi}P_{1\leftrightarrow 2}(\omega)$). In many-body systems this approach can be generalized to take into account all multiscattering processes~\cite{BenAbdallahEtAlMB2011,KruegerEtAl2012,LatellaEtAl2017,Latella2017,BiehsEtAl2021}. Formally the net power received by each object can be written in a Landauer-like form as
\begin{equation}
  \mathcal{P}_{i}=\underset{k\neq i}\sum\int_0^\infty\frac{d\omega}{2\pi}\hslash\omega[n(\omega,T_k) {\cal T}_{ki}(\omega)-n(\omega,T_i) {\cal T}_{ik}(\omega)],
\end{equation}
where the transmission functions ${\cal T}_{ik}$ are related to the coupling efficiency of modes at the frequency $\omega$ between the body $i$ and body $k$. Explicit expressions for many-particle systems within the dipole model and for multilayer systems can be found in Ref.~\cite{BenAbdallahEtAlMB2011} and Ref.~\cite{LatellaEtAl2017}, respectively, and general expressions derived within the scattering-matrix approach can be found in Ref.~\cite{KruegerEtAl2012}. For more details on the many-body theory and an extensive list of works on this topic we refer to the review~\cite{BiehsEtAl2021}. Generally the transmission functions depend on the geometric configuration and in particular on the distance between the objects $i$ and $k$ as well as on the optical material properties $\epsilon_{i,\mu \nu}$ and $\epsilon_{k,\mu \nu}$.  This opens up the possibility to tune the heat transfer by changing the configuration of the involved objects. More interesting, if the material properties significantly depend on temperature, external electric or magnetic fields, the heat flux can be actively controlled by changing these quantities.

\section{Rectification}
\label{sec:rectification}

\begin{figure}[H]
	\centering
	\includegraphics[width=0.8\textwidth]{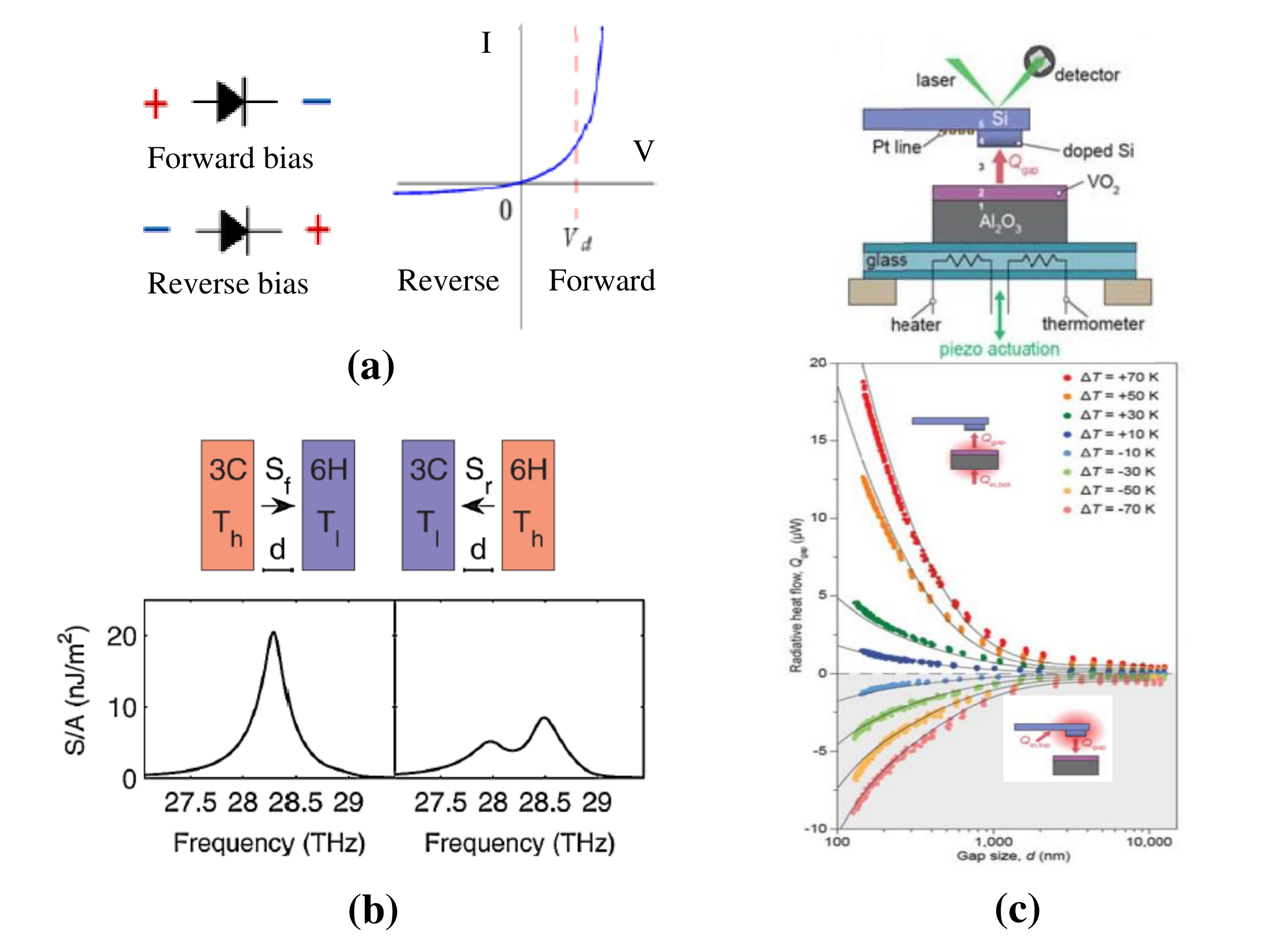}
      \includegraphics[width=0.5\textwidth]{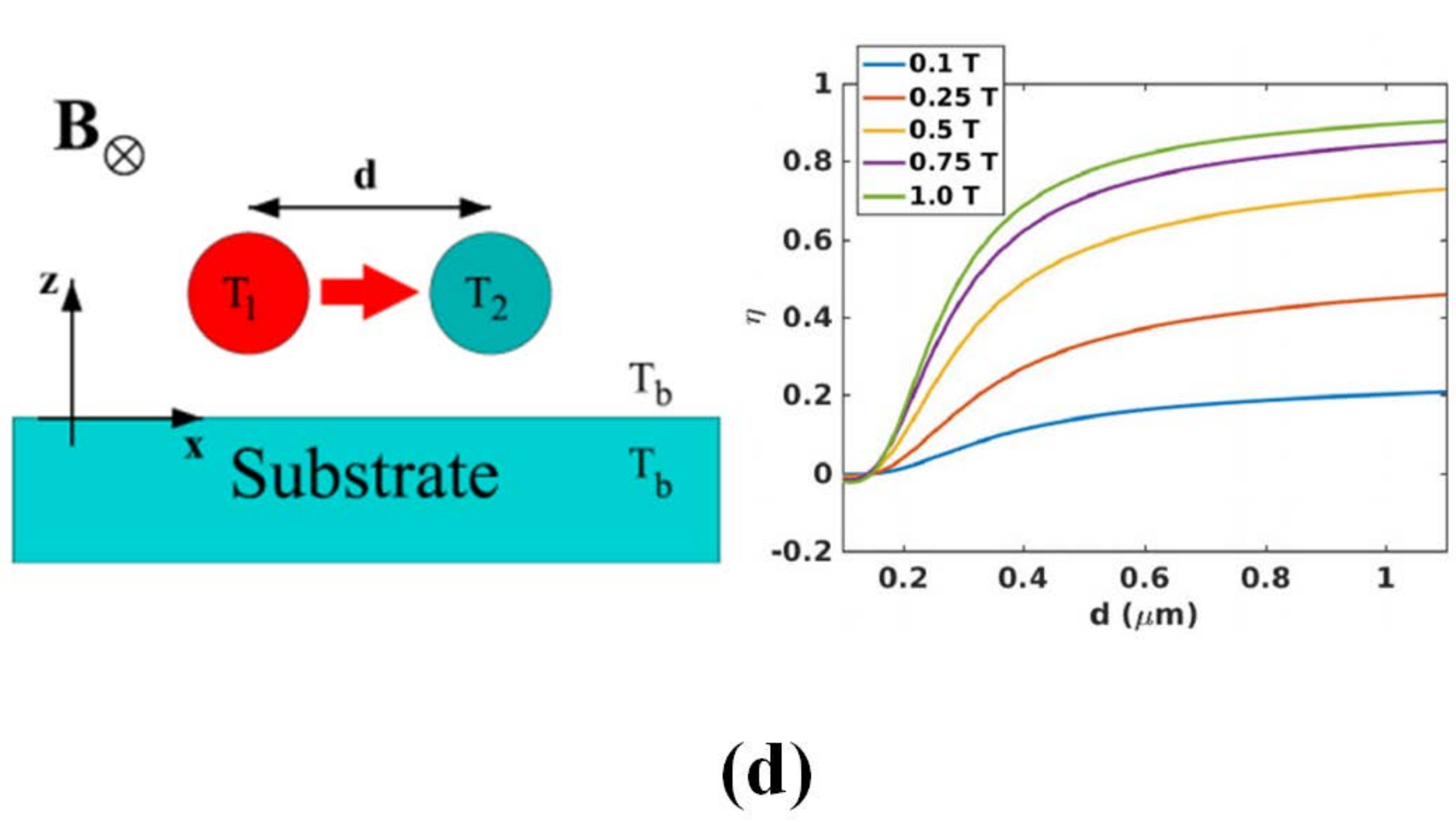}
	\caption{(a) Sketch of an electric diode and its typical current-voltage 
	curve. (b) Schematic of a near-field thermal rectifier made with a slab of $3C$-SiC and a slab of $6H$-SiC in the forward (f), and  the reverse (r) scenarios and the corresponding spectra of net heat flux. Reproduced with permission from~\cite{Ottey}. (c) Schematic of 
a microfabricated VO$_2$ based (phase-change material) radiative diode and measured near-field heat flux vs. the temperature bias $\Delta T$ in the forward ($\Delta T>0$) and the reverse  ($\Delta T<0$) scenarios. Reproduced with permission from~\cite{Fiorino}.
(d) Radiative thermal diode driven by nonreciprocal surface waves: the surface waves induce an asymmetry in the heat transfer between two magneto-optical nanoparticles placed close to a magneto-optical substrate in presence of a magnetic field $B$. This asymmetry is quantified by the rectification coefficient $\eta=(P_1-P_2)/P_1$, where $P_1$ is the power received by particle 1 when $T_2>T_1$ (backward scenario) while $P_2$ is the power received by particle 2 in the opposite situation (forward scenario). Reproduced with permission from~\cite{Ott}.} 
	\label{Fig2}
\end{figure}

The diode is one of the fundamental building blocks to control electron currents in electronic systems.
In an electronic diode (Fig~\ref{Fig2}-a), the current can flow mainly in one direction when a bias voltage is applied through its two terminals. This corresponds to a strong asymmetry due to a nonlinearity in the electrical conductance. 
As in the case of electronics, one of the basic devices to impose directionality of radiative heat flows are the thermal diodes. When a temperature bias $T_1-T_2$ is applied between two separated solids with temperatures $T_1$ and $T_2$, the magnitude of the heat flux they exchange by radiation generally does not depend on the sign of this bias. However, in presence of temperature-dependent material properties of the receiver or emitter, an asymmetry can appear between the heat flux $P^f$ in the forward biased situation ($T_1-T_2>0$) and the heat flux $P^r$ in the reverse scenario ($T_1-T_2<0$), such that $P^f \neq P^r$. Hence, radiative thermal rectification can be achieved under these conditions. Notice that here $P^f$ and $P^r$ are assumed positive, since they represent the heat flowing from the hottest to the coldest terminal in the two temperature biased scenarios.

As happens with the electronic counterparts, radiative thermal diodes act as a good radiative thermal conductor for a given sign of the temperature bias, while they behave as an insulator in the opposite situation. A first thermal radiative rectifier (Fig~\ref{Fig2}-b) has been introduced by Otey et al. in 2010~\cite{Ottey} using two different polytypes of SiC having different temperature-dependent optical properties. In this case, the transmission function ${\cal T}_{12}$ depends implicitly on the temperature of the two solids through the temperature dependence of their reflection coefficients $r_1$ and $r_2$. Thus, in the forward scenario with a low temperature $T$ and a high temperature $T+\Delta T$ we formally have a transmission function of the form
\begin{equation}
  {\cal T}^f_{12}={\cal T}_{12}(r_1(T+\Delta T), r_2(T)),
\end{equation}
while in the reverse scenario this function reads
\begin{equation}
  {\cal T}^r_{12}={\cal T}_{12}(r_1(T), r_2(T+\Delta T)).
\end{equation}
The heat transport asymmetry in the device can then be evaluated with the (normalized) rectification coefficient 
\begin{equation}
  R=\frac{| P^f- P^r | }{\max( P^f, P^r )}.
\end{equation}

When the interacting solids have weakly temperature dependent optical properties, $R$ is relatively small provided that the temperature bias is small as well. Hence, a rectification coefficient of ut to $\approx 29\%$ has been reported~\cite{Ottey} in near-field regime between two planar slabs of $3C$-SiC and $6H$-SiC with $\Delta T=300\,$K  or  between  slabs covered by an optimized coating  ~\cite{IizukaJAP2012} or slabs made with doped semiconductors with different doping levels and different thicknesses~\cite{Basu}. On the other hand, rectification coefficients as high as $90\%$ have been reported  between two solids when the temperature bias becomes large~\cite{Wang2013, XuEtAl2018, Zhu}.

In 2013, phase-change materials have been proposed~\cite{pba2013,YangEtAl2013,Huang} to improve the asymmetry of the radiative transport in configurations leading to large rectification coefficients with a relatively small temperature bias. These materials undergo a sudden and drastic change in their optical properties
around their critical temperature. Among these materials, metal-insulator transition (MIT) materials have attracted significant attention to design radiative heat rectifiers~\cite{pba2013,YangEtAl2013,Huang,Gu,Ghanekar,Zheng,Ghanekar2,Chen}. A widely MIT material is vanadium dioxide (VO$_2$)  which undergoes its phase transition at $T_c\approx340\,$K~\cite{Baker,Qazilbash}. Thanks to this transition rectification coefficients higher than $70\%$ have been predicted and highlighted in far-field regime~\cite{pba2013,Ito,Forero-Sandoval} with a temperature bias $\Delta T<50\,$K  and values around $90\%$ have been observed in near-field regime~\cite{Huang,ZwolEtAl2011,Zwol,Fiorino}. Furthermore, materials undergoing a normal-metal-superconductor transition~\cite{Kralik2017,Musilova2019,NefzaouiEtAl2014,OrdonezJAP2017,CuevasPRA2021} have also been considered to design radiative thermal rectifiers operating at cryogenic temperatures with similarly good performances.

Non-reciprocal  materials have also been considered to break the symmetry in the heat transport berween two bodies. Rectification factors close to $90\%$ have been recently predicted~\cite{Ott} in systems made with magneto-optical (MO) particles placed above a MO substrate (Fig.~\ref{Fig2}-d.) and exchanging heat via surface waves.

Finally, a concept of many-body rectifier working by embedding a passive intermediate body interacting with the two terminals, has been recently introduced~\cite{LatellaarXiv2021}. Unlike  the classical thermal rectification discussed above which require a noticeable temperature dependence of the optical properties of the materials, here the asymmetry in the heat transport results only from many-body interactions. Hence they can rectify the heat flux  over a broad temperature range. 

\section{Modulation and switching}

Controlling the magnitude and the direction of heat flux exchanged between solids at nanoscale is of prime importance in many technological applications and considerable effort has been made these last years to develop new strategies to this end. Below we discuss the recent developments carried out in this direction. 
\begin{figure}[H]
	\centering
	\includegraphics[width=0.8\textwidth]{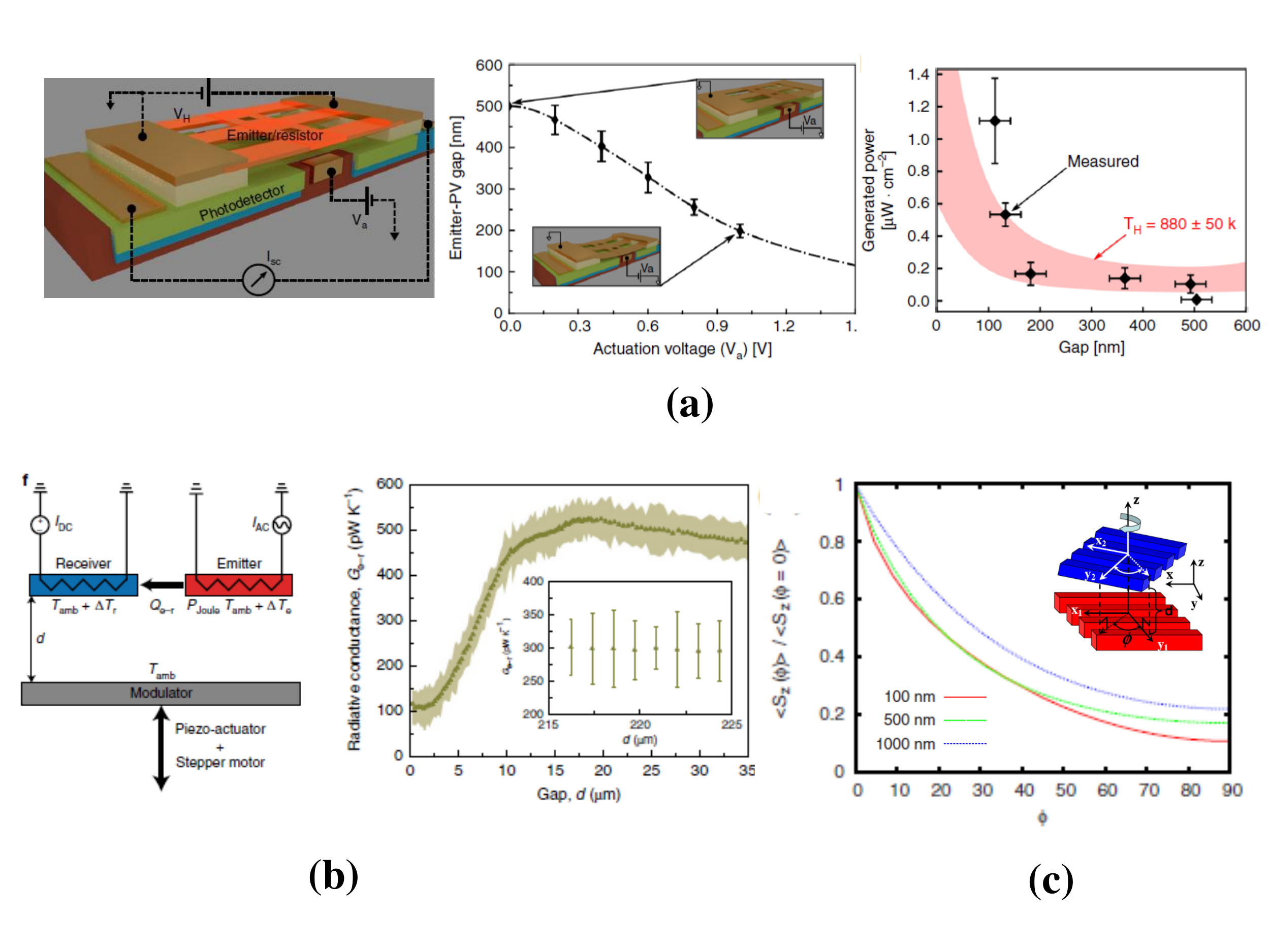}
       \includegraphics[width=0.55\textwidth]{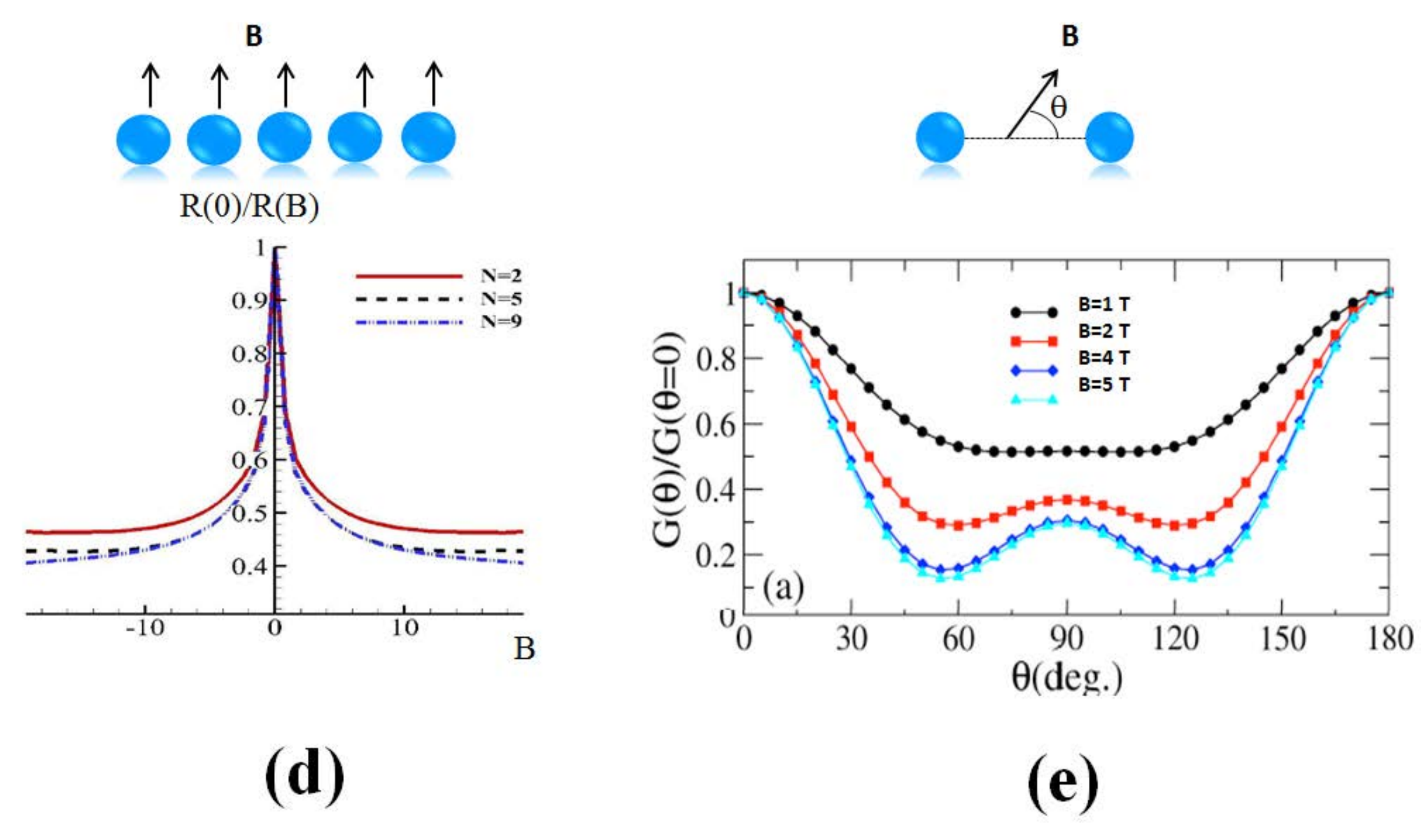}
	\caption{(a) Nano-electromechanical thermal switch. The NEMS consists in 
                 a suspended bridge (thermal emitter) which is brought closer to a solid through application of an actuation potential $V_a$. 
		 Their separation distance and the heat flux they exchange can be controlled with $V_a$. Reproduced with permission from~\cite{Lipson2}.
                 (b) Active control of heat flux exchanged between two solids. The thermal 
                 conductance between the emitter and the receiver is changed by adjusting the distance which separate them from a third body 
		 using a piezoelectric actuator. Reproduced with permission from~\cite{Thompson}. (c) Heat flux exchanged in near-field regime between two twisted gratings. Reproduced with permission from~\cite{Biehs2011}.
(d) Giant thermal magnetoresistance in plasmonic structures: the thermal magnetoresistance of magneto-optical nanoparticle chains changes drastically with respect to the strength of an external magnetic field $B$ orthogonal to the chain. Reproduced with permission from~\cite{Latella2017}.
	(e) Anistropic magnetoresistance: the thermal conductance between two magneto-optical particles changes with respect to the orientation of an applied magnetic field. Reproduced with permission from~\cite{Ekeroth}.} 
	\label{mechanical}
\end{figure}

\begin{figure}[H]
	\centering
	\includegraphics[width=0.8\textwidth]{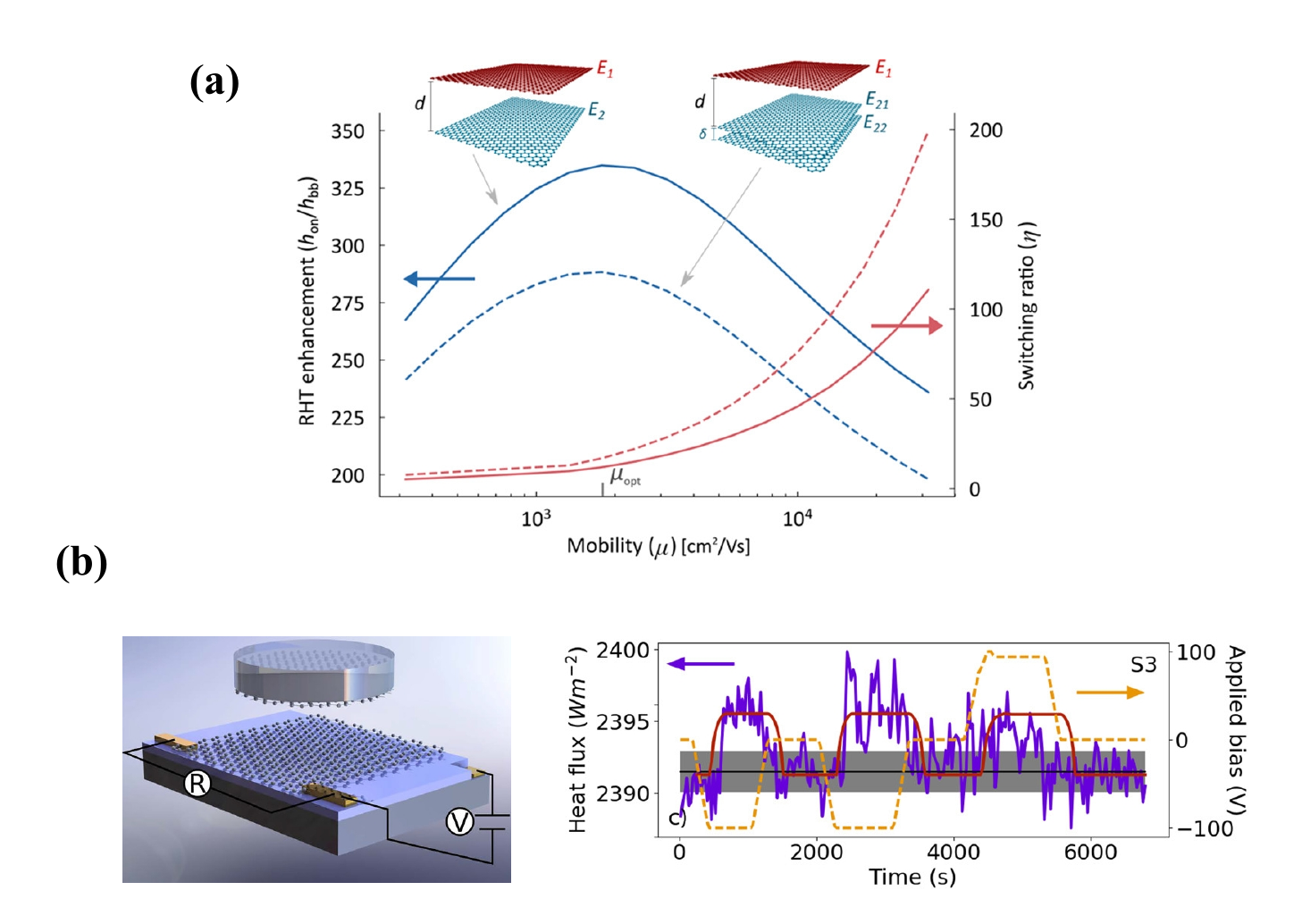}
	\caption{(a) Radiative heat transfer enhancement (normalized to the blackbody heat transfer coefficient) and heat flux switching  between two graphene sheets or between a single sheet and a stack of graphene as a function of the carrier mobility. Reproduced with permission from~\cite{IlicEtAl2018}. (b) Measured heat flux and applied bias as function time between a graphene-coated silica optical flat and a backgated, graphene-coated silicon wafer with a gate dielectric of SiO$_2$ and Al$_2$O$_3$ ($285\,$nm and $8\,$nm, respectively). Due to a non-negligible thermal capacitance, a time delay (about 3 min) from when the bias is
applied and when the heat flux change is observed can be observed. Reproduced with permission from~\cite{ThomasEtAl2019}.} 
	\label{graphene}
\end{figure}

The most natural way to control the magnitude of flux  exchanged between two solids is by changing their separation distance $d$.  In near-field regime the transmission coefficient scales like $\mathcal{T}_{12} \propto 1/d^n$, where $n = 6$ for two nanoparticles, $n = 3$ for a spherical object in vicinity of a slab, and $n = 2$ for two slabs, for instance. It follows that a displacement of one decade from a given position modifies the heat flux by orders of magnitude. This property can be exploited by mechanically changing the separation distance between two objects to modulate or switch the near-field radiative heat flux. 
Furthermore, micro/nano electromechanical systems (MEMS/NEMS) have been developped~\cite{Lipson, Lipson2,LimEtAl2013} in the last years which allow for performing a high precision control of the separation distance between two solids in the subwavelength regime, up to distances of few tens of nanometers, by tuning the electrostatic interaction between the solids using an actuation potential (Fig.\ref{mechanical}-a).  MEMS technology can be used for an active thermal management at nanoscale or to harvest on demand the near-field energy confined at the surface of hot objects using tunable near-field thermophotovoltaic converters~\cite{Lipson2}. Besides this control of near-field heat exchanges through a change of the separation distance between the solids, multiscattering effects induced by the presence of a third body has been proposed~\cite{Thompson} to tune the  heat exchanges in the near-field between two solids (Fig.\ref{mechanical}-b). By bringing a third body (even non-emitting) close to the emitter and receiver the heat flux exchanged between these bodies can be either amplified or inhibited thanks to many-body interactions.

Another way to control mechanically the near-field heat exchanges between two solids is the change of their relative orientation keeping their separation distance constant. This can have a strong impact in the coupling effciency of evanescent modes supported by each solid and therefore on the heat flux they exchange. Such change can simply be achieved using textured solids in relative rotation. For instance, the radiative heat flux between two uniaxial slabs with optical axis within the interface can be tuned by relative rotation of one slab~\cite{Biehs2011} as shown for two grating structures in Fig.\ref{mechanical}-c. When the optical axes are aligned, the heat flux is maximal and when the optical axis are perpendicular to each other the heat flux has a minimum. This effect can of course be exploited for any two anisotropic media, and it has already been demonstrated~\cite{Wu2021b} for two natural hyperbolic materials like hexagonal Boron Nitride (hBN). It must be noted that the operating speed of these mechanical control is intrinsically limited by the thermalization time of its components. With nanostructures interacting in near-field this time is typically in the order of few milliseconds.
Moreover, it is worth to point out that mechanical controlled actuation may be difficult to implement in certain situations and moving parts in any device are usually not desirable because of wear and tear. 
Beside the mechanical control of the separation distance and relative orientation strain-controled switches have been recently proposed to tune the flux. In these systems an intermediate layer of material, whose permittivity is controlled with mechanical strain, drives the radiative heat flux between a source and a drain at fixed separation distances~\cite{Papadakis2}.

As shown in the previous section MO materials can also be used to control
actively the near-field heat exchanges between two solids using an external magnetic
field. This possibility has been first suggested by Moncada-Villa et al. ~\cite{Cuevas2015,MoncadaPRB2020} who have shown that a change of the magnitude of magnetic field can significantly modify both the nature and the coupling of evenescent modes. 
More recently new thermomagnetic effects in MO systems~\cite{Latella2017,Ekeroth} have opened the way to a new strategy for controlling near-field heat exchanges. The first effect is a giant magneto-resistance~\cite{Latella2017} which enables a
significant increase of the thermal resistance along MO nanoparticle networks (Fig.\ref{mechanical}-d)) with increasing magnitude of an external magnetic field. This giant resistance results from a strong spectral shift of localized surface waves supported by the
particles under the action of a magnetic field. 
Recent works have combined MO materials and dielectrics in a hyperbolic multilayer structure~\cite{Ekeroth,MoncadaPRB2021}, because on the one hand the formation of hyperbolic bands can increase the near-field radiative heat flux in such systems~\cite{BiehsPRL2012,GuoAPL2012,IizukaPRL2018} and on the other hand the application of a magnetic field enables a significant active modulation of the heat flux. However, it seems that the effective medium calculations in [57] predict an increasement of  the near-field heat flux for extremely large magnetic fields, whereas the exact calculations in~\cite{MoncadaPRB2021} predict a heat flux reduction for moderately large magnetic fields.

An alternative to such magneto-optical control is the electrical actuation of optical properties of materials. Among all materials, graphene-based materials~\cite{Novoselov,Geim}, have  shown to be good candidates to ensure this control. 
By changing the Fermi level of a graphene sheet deposited on a solid using an  external gating, the scattering properties of this solid can be actively modulated~\cite{Volokitin2011,Svetovoy2012,IlicEtAl2012,Dalvit2015,vanZwol2012,YangEtAl2018,IlicOE,RM2013,YangWang,IlicEtAl2018,ThomasEtAl2019,He2,He3,LimEtAl2013,RMEtAl2017}. This electrical actuation of optical properties of graphene-based materials has been exploited to efficiently tune and even amplify the near-field heat exchanges between two solids (see Fig.~\ref{graphene}). 
The ferroelectrics state of some materials can also be tuned to control the radiative heat exchanges~\cite{HuangEtAl2014}. The active change of their spontaneous polarization can be used to shift the resonance frequency of surface phonon-polariton which some of these materials support and consequently control radiative heat transfer via varying external electric fields. 
Recently, three-body systems made with graphene-based  materials coupled with  ferroelectrics have demonstrated their strong potential to modulate near-field heat flux at kHz frequencies~\cite{LatellaarXiv2021_2}.
Finally switching and modulation of heat flux has been highlighted using metal–oxide–semiconductors (MOS).  Analogously to the MOS capacitor in electronics, the accumulation and depletion of charge carriers in an ultrathin plasmonic film can be used to control the coupling of surface waves~\cite{PapadakisACS2019}. 

\begin{figure}
	\centering
	\includegraphics[width=0.8\textwidth]{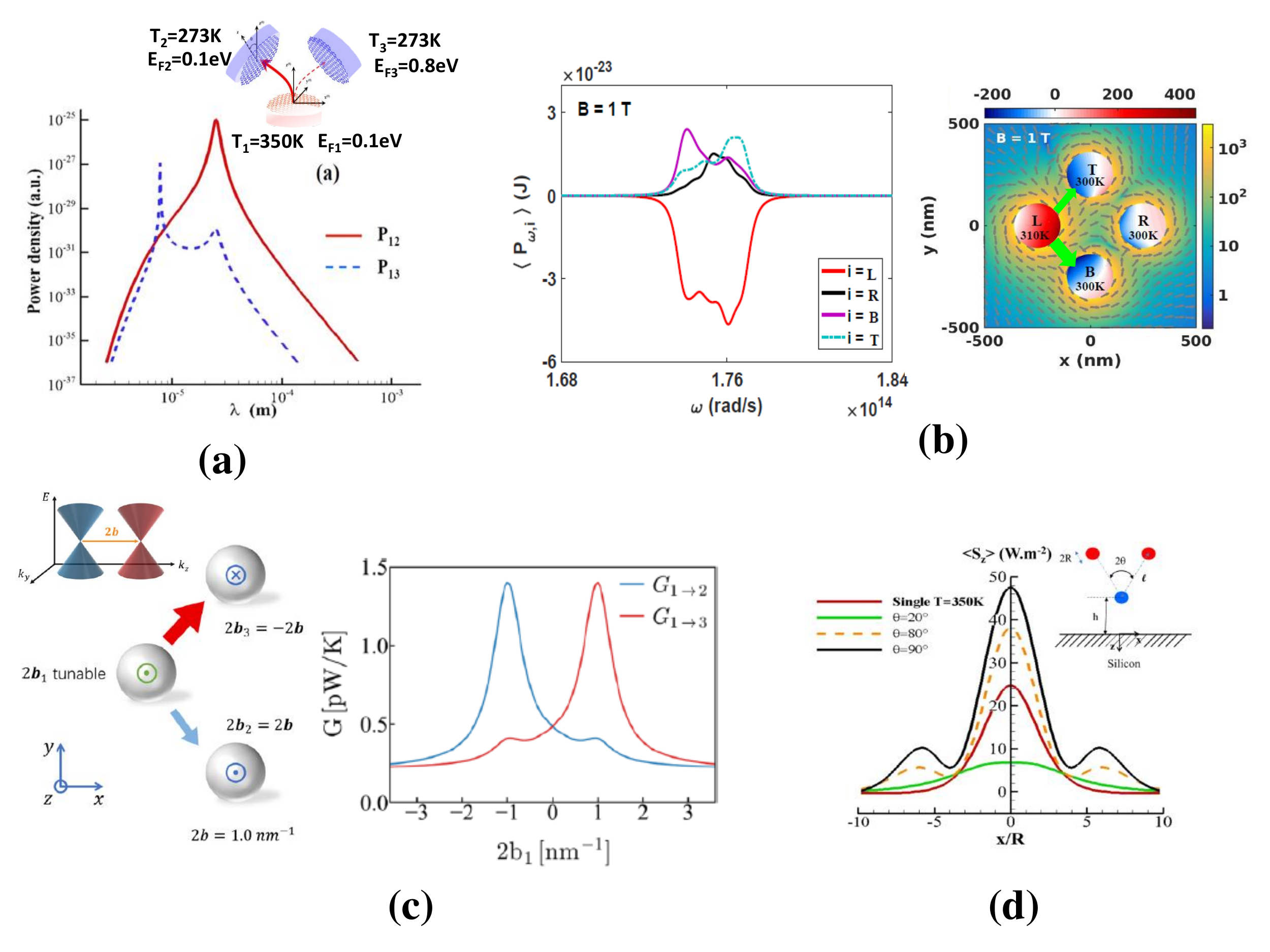}
	\caption{(a) Graphene-based heat flux splitter. The thermal powers $P_{12}$ and  $P_{13}$ exchanged in near-field regime between three identical pellets arranged in a symmetrical geometric configuration can be controlled by tuning the Fermi levels of graphene flakes deposited on their surface. Reproduced with permission from~\cite{pba2015}. (b) Spectral power and heat flux lines by radiative Hall effect in a four terminal junction made of MO particles forming a square. The junction is exposed to an external magnetic field $B$ in the direction orthogonal to the particle plane while $T_L=310\:K$ and $T_R=T_T=T_B=300\:K$. The mapping shows the Poynting vector field around the particles and illustrates the symmetry breaking  induced by the magnetic field. Reproduced with permission from~\cite{Ott2019}. (c) Radiative thermal router consisting of three spheres of the same radius made of magnetic Weyl semimetals forming an isosceles triangle in the $x-y$ plane. By tuning the Weyl node separation $2b_1$ of the first sphere located at the apex of this triangle the thermal conductances $G_{1\rightarrow 2}$ and $G_{1\rightarrow 3}$ can be controlled in an asymmetric way. Reproduced with permission from~\cite{Guo}. (d)  Heat flux focusing with a multi-tip SThM platform with three tips. The tip temperatures and their location are individually controlled, so that the thermal energy they radiate can be focused and even amplified in spots that are much smaller than those obtained with a single thermal source. Reproduced with permission from~\cite{pbaPRL2019}.} 
	\label{splitter}
\end{figure}

\begin{figure}
	\centering
	\includegraphics[width=0.8\textwidth]{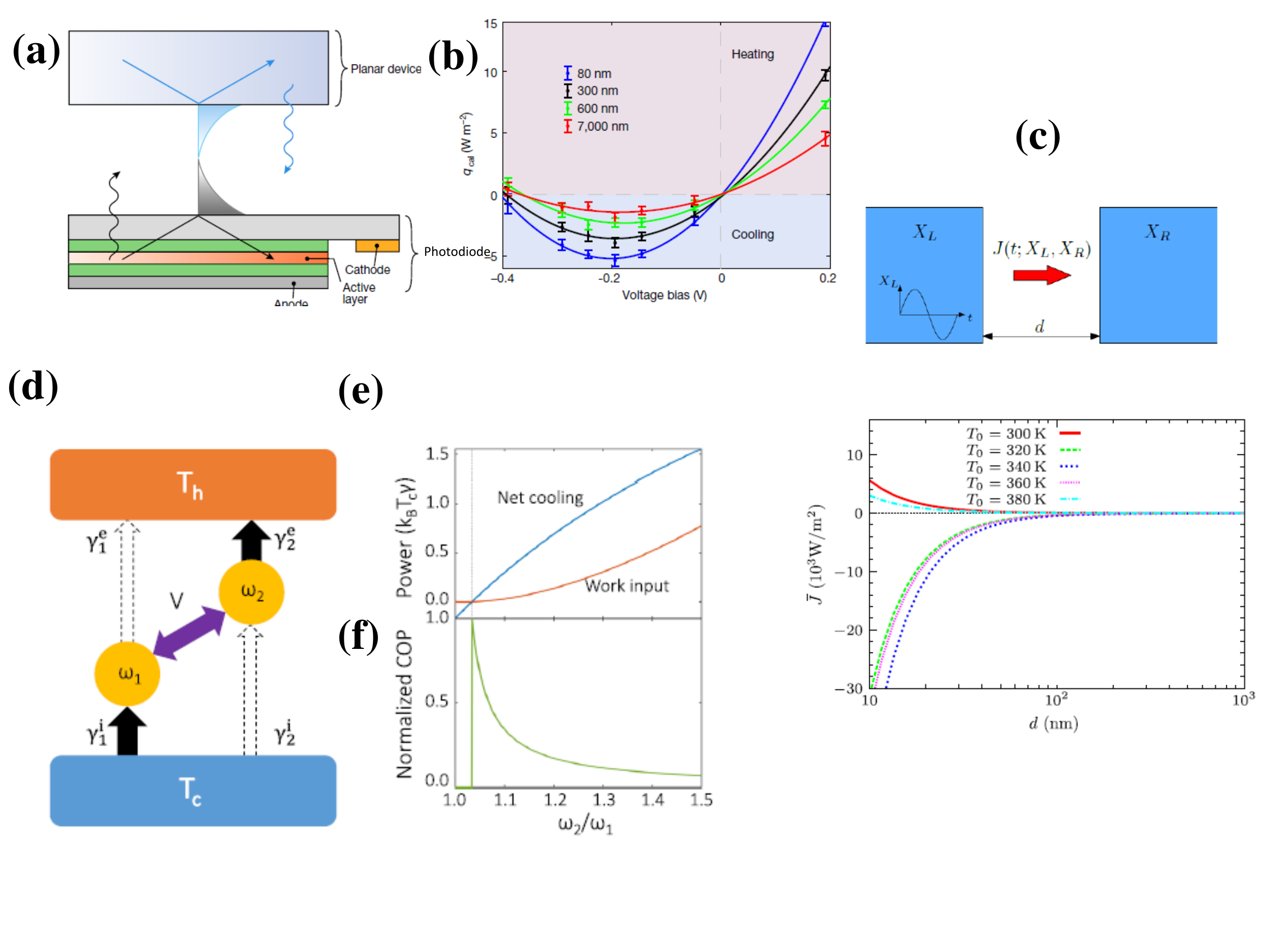}
	\caption{(a) Photonic refrigerator working in near-field regime. By applying a bias voltage on a photodiode its apparent temperature can be reduced, so heat can be extracted from a hot solid by radiation. (b) Heat flux exchanged in the photonic refrigerator with respect with the bias voltage. Reproduced with permission from~\cite{Zhu2019}. (c) Cooling by radiative heat shutting. By adiabatically modulating the temperature or the chemical potential of a solid, an extra flux superimposes to the steady state flux. Here we show the time-averaged flux $ \bar{ J } $ between a VO$_2$ slab and a sample of SiO$_2$ when the temperature of the VO$_2$ slab is $ T_L ( t ) = T_0 + \delta T \sin (\Omega t ) $ with an amplitude $ \delta T = 30\,$K, whereas the temperature of the other body is fixed at $ T_R = T_0 $. For a temperature modulation around the critical temperature of VO$_2$, the average flux can extract heat from the body with stationary temperature. Reproduced with permission from~\cite{Latella2018}. (d) Thermal photonic refrigerator operating between a cold solid at $T_c$ and a hot solid at $T_h$. Two modes at frequencies $\omega_{1}$ and $\omega_{2}$ are coupled through a time-modulation of the refractive index at a time scale faster than the thermal relaxation process. (e) Net cooling power and work input as a function of the ratio of the frequencies of the two modes, for $T_h=300\:$K and $T_c =290\:$K.  (f) Coefficient of performance (COP) of the refrigerator normalized to the Carnot limit. Reproduced with permission from~\cite{Fan_refrigeration}.} 
	\label{cooling}
\end{figure}

\section{Heat splitting and focusing}

The directional control of radiative heat flux exchanged in near-field regime in a set of solids can be achieved using various of the before mentioned mechanisms in order to break the symmetry. Hence, a heat flux splitting can be realized inside a set of pellets covered by graphene flakes by electrically tuning the Fermi level of graphene as sketched in Fig.\ref{splitter}-a~\cite{pba2015}. Such a control allows us to promote certain near-field interactions by tuning the graphene plasmons supported by the flakes.

The direction of heat flux can also be modified in magneto-optical systems using an external magnetic field. Indeed, as illustrated in Fig.~\ref{splitter}-b~\cite{Ott2019} in four terminal junction forming a square with $C4$ symmetry, when a temperature difference $\Delta T=T_L-T_R$ is applied between the particles $L$ and $R$  a radiative 
thermal Hall effect~\cite{pbaHall} transfers heat transversally to the primary gradient bending so the overall flux. This effect results from the fact that the transmission coefficients $\mathcal{T}_{ij}$ and $\mathcal{T}_{ji}$ are not equal  in nonreciprocal systems. 

Thermal routers~\cite{Guo} based on magnetic Weyl semimetals have been recently introduced using the unique properties of optical gyrotropy.
In these systems (Fig. \ref{splitter}-c), which consists of three spheres made of magnetic Weyl semimetals, the heat flux direction can be controlled by moving the Weyl nodes in the material using an external (magnetic or electric) field. It has also been shown that an anomalous photon thermal Hall effect can be realized in Weyl semimetals~\cite{OttEtAL2020}.

Recently, the concept of multitip scanning thermal microscopy (SThM) has 
been proposed~\cite{pbaPRL2019} to  locally focus and amplify the heat flux in regions much smaller than the diffraction limit and even smaller than the spot heated by a single tip. As illustrated in Fig.\ref{splitter}-d, the full width at half maximum (FWHM) of the spatial distribution of heat flux on the surface of substrate
can be significantly reduced in comparison with that with a single tip. For specific geometric configurations, the heat flux can even locally back 
propagate towards the emitting system which acts in this case as a heat pump.

\section{Active insulation, cooling and refrigeration}

While numerous research works have been devoted to the development of nanophotonic structures to control the far-field heat exchange and enable new applications in the field of radiative cooling, little attention has been paid so far to radiative cooling at subwavelength scales. During the last years some progress have been made in this direction and new mechanisms have been proposed to actively cool down solids through near-field heat exhange. 
The first advance in the development of solid-state photonic cooling operating in near field has been performed in 2015~\cite{Fan_cooling1}. The basic idea for this cooling mechanism consists in the use of a photodiode as illustrated in Fig.\ref{cooling}-a which is brought close to the solid to be cooled down. By applying an external bias voltage on the photodiode, photons are emitted with a non-vanishing chemical potential which follows from a spectral shift in the Bose-Einstein distribution function. Consequently, the apparent temperature of the photodiode can be artificially made smaller than its real temperature, so that heat can flow in the opposite direction of temperature gradient (Fig.~\ref{cooling}-b). Work is being performed on the photodiode, so there is no violation of any fundamental law. Moreover, the magnitude of heat exchange in the near field leads to a thermodynamic efficiency for such solid-state cooling device which is close to the Carnot limit. A proof-of-principle of this cooling principle has been demonstrated recently~\cite{Zhu2019}. 

\begin{figure}
	\centering
	\includegraphics[width=0.8\textwidth]{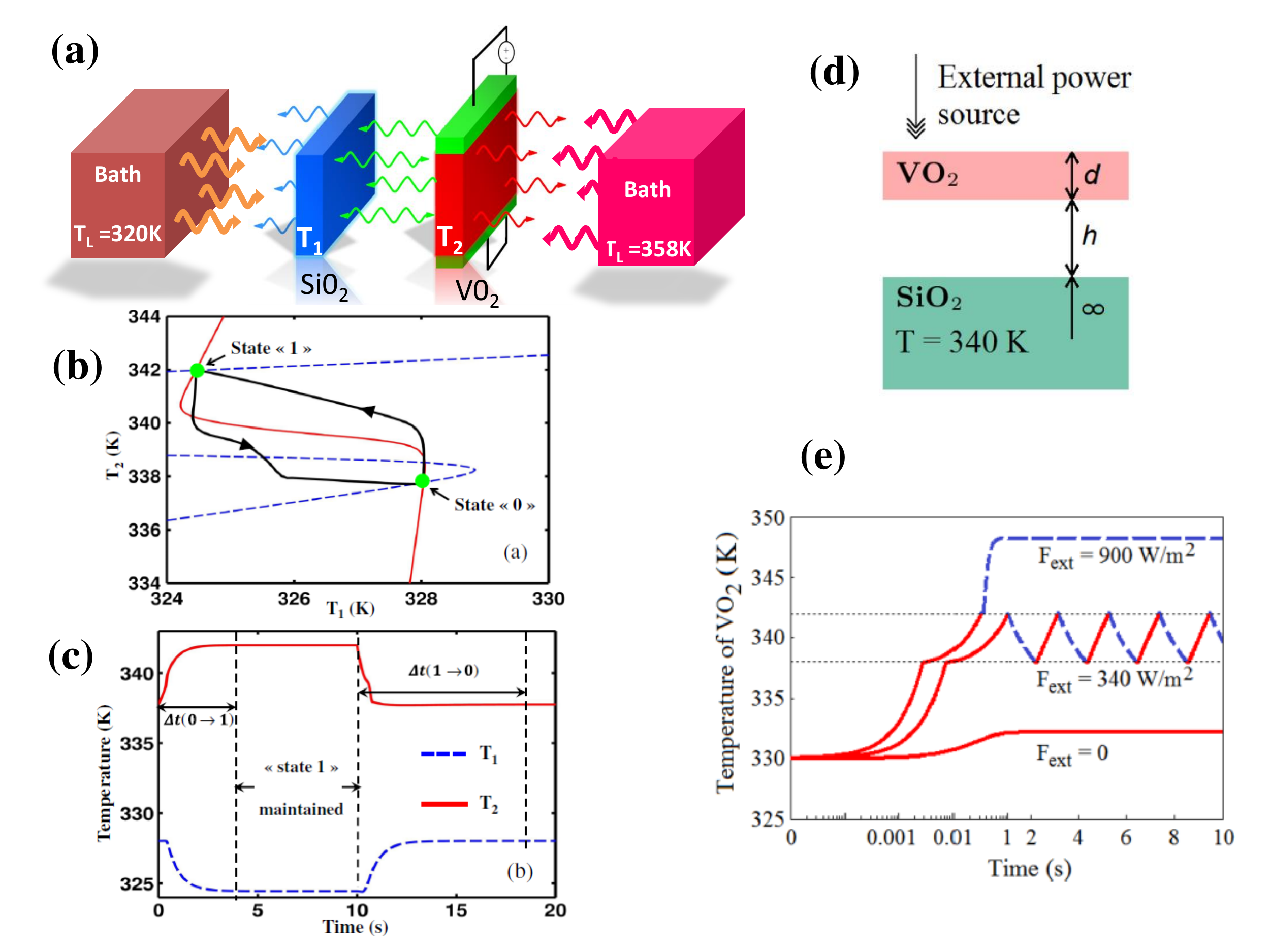}
	\caption{(a) Sketch of a (bistable) radiative thermal memory. A membrane 
made of  a phase transition material (VO$_2$) is placed at subwavelength distance from a dielectric layer (SiO$_2$). The system is surrounded by two thermal baths at different temperatures $T_L$ and $T_R$. The temperature of the VO$_2$ membrane can be increased or reduced either by Joule heating or by using Peltier elements. (b) Evolution of temperatures (black line)  between the two stable equilibrium temperatures (i.e. states ``$0$'' and ``$1$''). The dashed blue and solid red lines represent the local equilibrium conditions 
(vanishing flux) for each membrane. (c) Time evolution of SiO$_2$ and VO$_2$ membrane temperatures. The thermal states ``$0$'' and ``$1$'' can be maintained for an arbitrarily long time, provided no external heat flux perturbs the system. Reproduced with permission from~\cite{Kubytskyi}. (d) Self-thermal oscillator made with a VO$_2$ membrane in the vicinity of a SiO$_2$ substrate in presence of an external constant power source  $F_\mathrm{ext}$. (e) Time evolution of the VO$_2$ membrane temperature at different external powers. Reproduced with permission from~\cite{Dyakov}.} 
	\label{memory}
\end{figure} 

\begin{figure}
	\centering
	\includegraphics[width=0.8\textwidth]{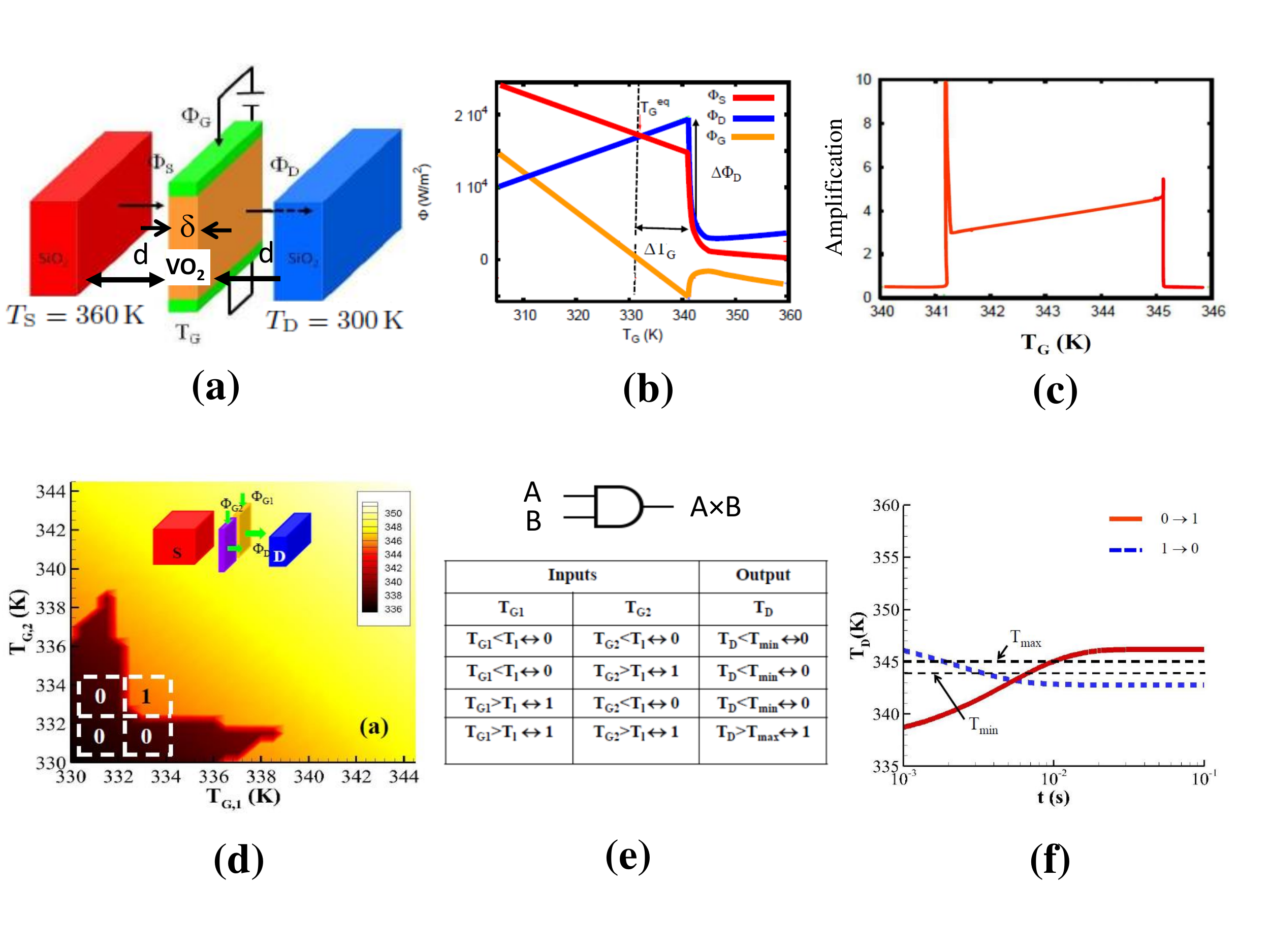}
	\caption{(a) Radiative thermal transistor made of a three-terminal system composed of a SiO$_2$ source, a VO$_2$ gate and a SiO$_2$ drain. The temperature of the VO$_2$ gate can be actively controlled around its critical temperature by an external thermostat. (b) Radiative fluxes $\Phi_S, \Phi_D$, and $\Phi_G$ exchanged between the different parts inside the transistor  when the source and the drain temperatures are fixed at  $T_S = 
360\,{\rm K}$ and $T_D = {\rm 300\, K}$, respectively. (c) Amplification factor with respect to the gate temperature. Reproduced with permission from~\cite{pba2014}. (d) Radiative AND gate realized by a double SiO$_2$ gate thermal transistor, the source being made of SiO$_2$ and the drain of VO$_2$. The source (S) is thick (no size effect), while both gates
(G1, G2) and the drain (D) have a thickness of $250$ and $500\:$nm, respectively. The separation distances are $d = 100\:$nm and the gates
are assumed  to be isolated one from the other. The color map represents the output $T_D$ with respect to the two input temperatures $T_{G1}$ and
	$T_{G2}$, the temperatures of two gates. The operating range of the AND gate is delimited by a dashed rectangular domain centered at ($T_{G1},T_{G2}$) = ($T_l,T_l $) with $T_l = 332.2\:$K. (e) Truth table for the ideal AND gate. (f) Typical relaxation dynamics in a logic gate. Reproduced with permission from~\cite{pba2016}.} 
	\label{transistor}
\end{figure} 

The active modulation of physical properties or intensive quantities has also been proposed to cool down solids through near-field interactions in 
two and many-body systems. Latella et al.~\cite{Latella2018} have considered radiative thermal exchange between two bodies, where the temperature of at least one body is adiabatically (slowly) modulated through interactions with external thermostats. Due to the nonlinear dependence of the temperature in the radiative heat exchange, the time average heat flux can proceed against the average temperature bias, even though instantaneously heat flows always from the hotter to the colder body.
When the modulation is performed in such a way that the number of modes which participate to the transfer decreases with the temperature, a radiative shuttling effect~\cite{Li2009} can dynamically pump heat from a body with stationary temperature in spite of a vanishing average thermal bias (Fig.\ref{cooling}-c). This situation can occur, for instance, in systems made with phase-change materials whose optical properties drastically change across the phase transition. Similar heat-pumping mechanisms driven by combined modulations of
positions and temperatures have been recently highlighted~\cite{Messina2020} in many-body systems.

Photonic refrigeration can also be observed in systems whose refractive index undergo a temporal modulation~\cite{Fan_refrigeration}. In these systems (Fig.\ref{cooling}-d), two resonant modes such as cavities modes inside the solid to be cooled down are coupled and driven by a time modulation of refractive index. When this modulation is turned on, a fraction of the thermally generated photons from the mode of lowest energy are up-converted to the second mode and emitted in the surrounding environment. These photons carry a power (Fig.\ref{cooling}-e) away from the solid with a high coefficient of performance (Fig.\ref{cooling}-f).

\section{Logical circuits}

Besides the control of heat fluxes, thermal information processing at the nanoscale remains today a challenging problem. Some building blocks have been introduced during the past years in this regard, with the aim of establishing thermal analogues of conventional electronic building blocks which are driven by thermal photons rather than by electrons. Among these devices, 
multistable systems have been proposed to store the radiative energy~\cite{Kubytskyi} and to release it into the environment upon request. For example, systems composed of phase-change materials have several  equilibrium (stationary) temperatures and behave like thermal memories. As shown in (Fig.\ref{memory}-a) in the particular case of a bistable system~\cite{Chinmay} which consists in two slabs of temperature $T_1$ and $T_2$, which mutually interact and which are coupled to two thermal reservoirs, two stable equilibrium temperature can exist. These states "$0$'' and "$1$'' correspond to the temperature pair $(T_1, T_2)$ for which the heat flux received by each slab vanishes (Fig.\ref{memory}-b). Such states can be maintained for arbitrarily long times (Fig.\ref{memory}-c), provided that the temperatures of reservoirs are kept constant and no external perturbation modifies the net flux on each slab. By heating or cooling the slab made with the phase-change material, the thermal state of the system switch from one state to the other.
This switching has been used to design self-induced thermal oscillators~\cite{Dyakov} by exploiting the hysteretic behavior of the phase-change material around its critical temperature (Fig.\ref{memory}-d,e).

Another building block is the transistor. In electronics this device is a key element which allows for switching but above all
for amplifying an electric current flowing through a solid using a simple external bias voltage. This building block is at the origin of modern electronics which have revolutionized our current life. In 2014, a radiative thermal analogue of a transistor has been introduced~\cite{pba2014}. As its electronic counterpart, the radiative transistor is a three-terminal system (Fig.\ref{transistor}-a) composed by a hot body (the source), a cold body (the drain) and an intermediate slab made of a phase-change material (the gate). By operating at temperatures close to the critical temperature where the phase transition takes place in these materials, the heat flux received by the drain can be switched (Fig.\ref{transistor}-b), modulated and even amplified (Fig.\ref{transistor}-c) with a weak variation of the gate temperature. This behavior is closely related to the strong change in the optical properties of the phase-change material around its critical temperature. In this temperature range, the thermal resistance $R=(\frac{\partial \varphi_D}{\partial T_G})^{-1}$ defined as the variation of flux  $\varphi_D$ received by the drain with respect to the gate temperature $T_G$ is negative~\cite{Zhu2}. Under these conditions, the amplification factor of the 
transistor $A=\mid\frac{\partial \varphi_D}{\partial \varphi_G}\mid$ can be higher than unity~\cite{pba2014,CuevasPRA2021}.

By using single or combining several radiative transistors, logic gates have been designed~\cite{pba2016,Kathmann} to perform a Boolean treatment of information with heat exchanged in near-field regime. In Fig.\ref{transistor}-d we show an example of an AND-like gate made with a double gate transistor, where the gates are made of sililica and the drain is made of a phase-change material (VO$_2$). In this system, the temperatures $T_{G1}$ and $T_{G2}$ of the gates set the two inputs of the logic gate, and the temperature $T_D$ of the drain stands for the logic gate output. By introducing a threshold value for $T_D$ beyond which the output state of the gate switches from state ``$0$'' to state ``$1$'', we see that the system behaves as a digital AND gate (Fig.\ref{transistor}-d,e). The overall operating time of the logic gate corresponds to the time required to switch from one state to the other. This time is directly related to the thermalization of each element through radiative interactions, and in nanostructured systems it is of the order of few milliseconds (Fig.\ref{transistor}-f).

\section{Outlook}
The spatio-temporal control of near-field radiative heat exchanges in complex solid architectures has opened the way to a new generation of devices for both a passive and active thermal management at nanoscale. The new degree of freedom enables the development of wireless sensors working with heat as primary source of energy rather than with electricity. In such devices, heat coming from various heat sources (machines, electric devices…) can be captured, stored in thermal blocks (thermal capacitors or thermal memory) and used to launch sequences of logical operations in order to either control the heat flux propagation (direction, magnitude), trigger specific actions (opto-thermo-mechanical coupling with MEMS/NEMS, ignitiate chemical reactions…) or even make information treatment with heat. In this perspective the operating speed of this technology could be a limiting factor. Indeed, in circuits involving interacting nanostructures the typical timescale to process one single operation is of the order of milliseconds or even more due to the thermal inertia of building blocks. For information processing this speed is obviously not competitive with the current electronic devices but it is more than enough for active thermal management and thermal sensing. For example, existing near-field probes like those developed in the last decades~\cite{Kittel2008,Huth2011,DeWilde,Jones,WengEtAl2018} can be further advanced to measure some of the theoretically proposed modulation effects locally, whereas new multi-tip or many-body setups like that in Ref.~\cite{Thompson} are necessary to realize some of the thermotronic building blocks like the transistor. Nevertheless important progress could be done by considering 2D materials or solids far from their equilibrium where the heat carriers have different temperatures. In this last case the operating speed of thermal circuits could be reduce to few microsecond or even picoseconds, the typical relaxation time of electrons in solids. But this ultrafast physics of heat exchanges remains today a challenging problem both on a fundamental and practical point of view.


\begin{acknowledgments}
P. B.-A. acknowledges financial support from the Agence Nationale de la Recherche in France through the ComputHeat project (ANR-19-MRSEI-6786), S.-A. B. acknowledges financial support from the Heisenberg Programme of the Deutsche Forschungsgemeinschaft (DFG, German Research Foundation) under the project (No. 404073166) and I.L. acknowledges financial support from the European Union’s Horizon 2020 research and innovation programme under the Marie Sklodowska-Curie grant agreement No 892718.
\end{acknowledgments}





\end{document}